\documentclass[twocolumn,amsmath,amssymb,aps,prb,superscriptaddress]{revtex4}
\usepackage{graphicx}
\usepackage{bm}
\usepackage[usenames]{color}
\begin{document}

\title{Neutron scattering study of magnetic ordering and excitations in the ternary rare-earth diborocarbide Ce$^{11}$B$_2$C$_2$}

\author{Isao Nakanowatari}
\affiliation{Neutron Science Laboratory, Institute for Solid State Physics, University of Tokyo, 106-1 Shirakata, Tokai, Ibaraki 319-1106, Japan}

\author{Rei Morinaga}
\affiliation{Neutron Science Laboratory, Institute for Solid State Physics, University of Tokyo, 106-1 Shirakata, Tokai, Ibaraki 319-1106, Japan}

\author{Takahiro Onimaru}
\altaffiliation[Present address: ]{Department of Quantum Matter, ADSM, Hiroshima University, 1-3-1 Kagamiyama, Higashi-Hiroshima 739-8530, Japan}
\affiliation{SORST, Japan Science and Technology Agency, Kawaguchi, Saitama, Japan}
\affiliation{Neutron Science Laboratory, Institute for Solid State Physics, University of Tokyo, 106-1 Shirakata, Tokai, Ibaraki 319-1106, Japan}

\author{Taku J Sato}
\email{taku@issp.u-tokyo.ac.jp}
\thanks{Corresponding author}
\affiliation{Neutron Science Laboratory, Institute for Solid State Physics, University of Tokyo, 106-1 Shirakata, Tokai, Ibaraki 319-1106, Japan}
\affiliation{SORST, Japan Science and Technology Agency, Kawaguchi, Saitama, Japan}

\date{\today}

\begin{abstract}

Neutron scattering experiments have been performed on the ternary rare-earth diborocarbide Ce$^{11}$B$_2$C$_2$.
The powder diffraction experiment confirms formation of a long-range magnetic order at $T_{\rm N} = 7.3$~K, where a sinusoidally modulated structure is realized with the modulation vector ${\bm q} = [0.167(3), 0.167(3), 0.114(3)]$.
Inelastic excitation spectra in the paramagnetic phase comprise significantly broad quasielastic and inelastic peaks centered at $\hbar \omega \approx 0, 8$ and 65~meV.
Crystalline-electric-field (CEF) analysis satisfactorily reproduces the observed spectra, confirming their CEF origin.
The broadness of the quasielastic peak indicates strong spin fluctuations due to coupling between localized $4f$ spins and conduction electrons in the paramagnetic phase.
A prominent feature is suppression of the quasielastic fluctuations, and concomitant growth of a sharp inelastic peak in a low energy region below $T_{\rm N}$.
This suggests dissociation of the conduction and localized $4f$ electrons on ordering, and contrasts the presently observed incommensurate phase with spin-density-wave order frequently seen in heavy fermion compounds, such as Ce(Ru$_{1-x}$La$_x$)$_2$Si$_2$.
\end{abstract}

\maketitle


\section{Introduction}
The ternary rare-earth diborocarbides REB$_2$C$_2$ (RE: rare-earths) with heavy RE elements attract special attentions recently, as they exhibit intriguing successive ordering at low temperatures.
Exemplified by DyB$_2$C$_2$, the specific heat measurement detects two anomalies at $T_{\rm c2} = 15.3$~K and $T_{\rm c1} = 24.7$~K.~\cite{yamauchi99}
The higher temperature anomaly at $T_{\rm c1}$ is attributed to ordering of quadrupole moments by various measurements such as the ultrasonic vibration measurement~\cite{nemoto03} and resonant X-ray diffraction.~\cite{hirota00,tanaka04}
It has been suggested that the quadrupolar moments originate from nearly degenerated two Kramers' doublets realized by pseudo-cubic local symmetry around the Dy$^{3+}$ ions.
On the other hand, the lower temperature anomaly at $T_{\rm c2}$ is due to magnetic (dipole) ordering.
The magnetic structure established below $T_{\rm c2}$ is non-trivial; a complicated multi-${\bm q}$ ferrimagnetic structure is realized with the four modulation vectors ${\bm q} = (1,0,0), (0, 1, 1/2), (0,0,0)$ and $(0,0,1/2)$.
Such a structure cannot be stabilized by usual exchange-type interactions (including RKKY-type interactions), and strong influence by the quadrupolar order has been inferred.~\cite{yamauchi99}
As exemplified above, interplay of dipolar and quadrupolar degrees of freedom is essential for the complex successive ordering in the REB$_2$C$_2$ compounds with heavy RE elements.

In contrast, REB$_2$C$_2$ with the light rare-earth element RE = Ce does not have quadrupolar degree of freedom; absence of the ground state degeneracy has been reported in the ultrasonic vibration measurement.~\cite{onodera00}
Nonetheless, CeB$_2$C$_2$ is of particular interest because influence of magnetic interactions to the complex successive ordering may be separately deduced from that of the quadrupolar interactions by comparing ordering behavior of non-quadrupolar CeB$_2$C$_2$ with those in the heavy RE systems.
Also, strong electron correlations originating from larger hybridization between $4f$ and conduction electrons may bring about intriguing low-temperature properties.

Earlier studies on CeB$_2$C$_2$ may be summarized as follows.
Electronic specific heat coefficient $\gamma$ was estimated in the paramagnetic temperature range as $\gamma = 98.7$~mJ/mol K$^2$.~\cite{hirota93}
Thus, CeB$_2$C$_2$ may be classified as a heavy-fermion system with moderate mass enhancement, a typical consequence of strong electron correlations in the $4f$ electron systems.
A sharp anomaly at $T_{\rm N} = 7.3$~K was observed in the specific heat and magnetization measurements,~\cite{onodera00} indicating that a certain magnetic order is established below this temperature.
The specific heat anomaly is followed by a broad shoulder around $T = 6.5$~K, of which the origin is not elucidated.
To date, only one attempt was made to solve the magnetic structure; neutron diffraction experiment was performed using a single-crystal sample, and succeeded in observing a magnetic Bragg reflection at ${\bm Q} = (\delta, \delta, \delta')$ with $\delta \approx 0.161$ and $\delta' \approx 0.100$.~\cite{ohoyama03}
On the basis of this observation, an incommensurately modulated structure was inferred in the ordered phase.
However, because of insufficient quality of the crystal, it was the single peak that the diffraction experiment could detect, and consequently, the magnetic structure of the ordered phase could not be determined.
Inelastic neutron scattering has also been carried out to determine the crystalline-electric-field (CEF) splitting of the Ce$^{3+}$ 4f levels;~\cite{hillier06} the inelastic spectrum was measured at 7~K for $-15 < \hbar\omega < 45$~meV.
However, because of the limited energy range and temperature point, it hardly provides details of the spin fluctuations and excitations in CeB$_2$C$_2$.

Knowledge on the low-temperature magnetic structure is essential for the understanding of the magnetic ordering in the CeB$_2$C$_2$, and its relation to REB$_2$C$_2$ with heavy RE elements.
The CEF splitting and spin fluctuation spectrum are also crucial information to elucidate its magnetic properties.
Hence, to address these issues we have undertaken low-temperature neutron powder diffraction and inelastic scattering experiments in the present work.
In the powder diffraction we have succeeded in obtaining intensity data for a number of magnetic Bragg reflections, whereas inelastic spectra in a wide energy range up to 80~meV were collected at several temperatures below 100~K.
These experiments enable us to unambiguously determine the magnetic structure in the ordered phase, as well as peak energies in the inelastic spectrum; it will be shown that the lowest-temperature magnetic structure is a sinusoidally modulated structure with the modulation vector ${\bm q} = [0.167(3), 0.167(3), 0.114(3)]$, whereas the inelastic scattering detects broad quasielastic and two inelastic (CEF) peaks around $\hbar \omega \approx 0, 8$ and 65~meV in the paramagnetic phase.
The quasielastic fluctuations are suppressed below $T_{\rm N}$, and a sharp inelastic peak emerges in the low energy region.

\section{Experimental details}
Polycrystalline samples of Ce$^{11}$B$_2$C$_2$ (about 10~grams in total) were prepared using an arc furnace under an Ar gas atmosphere.
Purity of the starting elements was 99.9~\% for Ce and C.
To avoid strong neutron absorption of natural boron, we used the isotope enriched $^{11}$B (99.53~\% enrichment).
Structural quality of the resulting polycrystalline samples was checked by the X-ray powder diffraction as well as the neutron diffraction.

The polycrystalline samples were crushed into powder and loaded in a thin Al sample cell for the neutron scattering experiments.
The cell was then set to a $^{4}$He closed-cycle refrigerator.
Neutron powder diffraction was performed using a powder-diffraction detector bank newly installed to the LAM-80ET inverted-geometry time-of-flight (TOF) spectrometer at the KENS spallation neutron source, KEK, Japan.
The diffraction detector bank has 34 detectors covering the scattering angle range of $138^{\circ} < 2\theta < 162^{\circ}$.
We used the incident neutrons in a wave-length range of $1.51 < \lambda < 8.40$~\AA, which corresponds to the $d$-range of $0.8 < d < 4.3$~\AA.
Obtained powder diffractograms were analyzed using the home-made general-purpose Rietveld analysis code {\footnotesize MSAS-TOF}.~\cite{sato07}
To perform profile fitting of highly asymmetric peaks at large time-of-flights, a special profile function is used:
\begin{widetext}
\begin{equation}
f_{\rm prof}(t')  = \left \{
\begin{array}{ll}
  \frac{2 a_1 \alpha_1 [(1 - \eta)f_{\rm G}(t') 
      + \eta f_{\rm L}(t')]}{a_1 \alpha_1 + 2 \alpha} \ \ \ \ \ \mbox{for $t' < 0$,}\\
  \frac{\alpha \alpha_1 \{ [2 a_1 + (1 - a_1 - a_2) \alpha_1^2 t'^2 ] \exp(-\alpha_1 t')
  + 8 a_2 \alpha_1 t' \exp(-2 \alpha_1 t') \}}{a_1 \alpha_1 + 2 \alpha}
  \ \ \ \ \ \mbox{for $t' > 0$.}
  \end{array}
\right. 
\end{equation}
\end{widetext}
In the above expression, $t' = t_{\rm TOF} - t_0$, where $t_0$ is a time-of-flight at which a Bragg reflection appears.
The two functions $f_{\rm G}(t')$ and $f_{\rm L}(t')$ are the Gaussian and Lorentzian functions defined by:
\begin{eqnarray}
  f_{\rm G}(t') &=& \frac{\exp(-t'^2/\gamma_{\rm G}^2)}{\sqrt{\pi} \gamma_{\rm G}}, \nonumber\\
  f_{\rm L}(t') &=& \frac{\gamma_{\rm L}}{\pi (\gamma_{\rm L}^2 + t'^2)}.
\end{eqnarray}
The profile for negative $t'$ is a modified version of the pseudo-Voigt function,~\cite{wertheim74,thompson87} allowing independent values for $\gamma_{\rm G}$ and $\gamma_{\rm L}$.
On the other hand, to simulate a long tail due to the solid methane moderator at the KENS spallation source, a set of exponentially decaying functions is assumed for $t' > 0$ as inferred from the Ikeda-Carpenter equation.~\cite{ikeda85}
The parameters are assumed to be linearly $t_0$ dependent in the present TOF range: $\eta = \eta_1 + \eta_2 t_0$, $\gamma_{\rm G} =  \gamma_{\rm G1} + \gamma_{\rm G2}t_0$, $\gamma_{\rm L} =  \gamma_{\rm L1} + \gamma_{\rm L2}t_0$, $a_1 =  a_{11} + a_{12} t_0$, $a_2 =  a_{21} + a_{22} t_0$, $\alpha_1 =  \alpha_{11} + \alpha_{12} t_0$, and $\alpha = (1-\eta)/(\sqrt{\pi}\gamma_{\rm G}) + \eta/(\pi \gamma_{\rm L})$.
Details of the new diffraction bank at LAM-80ET and the analysis code will be published elsewhere.~\cite{shibata07,sato07}

Neutron inelastic scattering experiment was performed using the LAM-D inverted-geometry time-of-flight spectrometer, also installed at the KENS spallation neutron source.~\cite{ino93}
Final energy was fixed to 4.59~meV using the pyrolytic graphite (PG) 002 reflections, whereas higher harmonic neutrons were eliminated by cooled Be filters.
There are four analyzers at the scattering angles $2\theta = \pm35^{\circ}$ and $\pm85^{\circ}$; we mainly show data taken with the lower angle detectors in this report, and higher angle data are used only for phonon subtraction purpose.
Energy resolution was estimated as $\Delta E = 0.42$~meV [full width at half maximum (FWHM)] at the elastic position using a vanadium standard.
Background was subtracted from the raw data using a proper combination of empty-cell and absorber runs, and absorption effect was corrected using numerically calculated absorption factors.

\section{Results and discussion}

\subsection{Crystal structure}

First of all, the validity of the newly installed detector bank and the home-made Rietveld analysis code is confirmed by solving the room temperature structure of Ce$^{11}$B$_2$C$_2$.
Figure~1 shows the powder diffraction pattern at the room temperature recorded without using the refrigerator.
The pattern is analyzed assuming two phases; one is the tetragonal CeB$_{2}$C$_{2}$ structure with the space group $P4/mbm$ determined by Onimaru {\it et al.},~\cite{onimaru99} whereas the other is polycrystalline Al used as the sample cell in the present experiment.
For both the phases, the lattice parameters, atom positions (except for the Al phase), isotropic atom displacement parameters ($B^{\rm iso}$), and preferred orientation parameters were optimized, in addition to the profile and background parameters.
Resulting calculated diffraction profile is shown in the figure by the solid line.
Difference between the observation and the calculation is also presented in the figure. 
Coincidence between the calculated and observed intensities is quite satisfactory.
Obtained parameters are listed in Table~I, which are in perfect agreement with the previous result.~\cite{onimaru99}
This confirms the validity of the newly installed detector bank and the Rietveld analysis code.

Figure~1(b) shows a powder diffractogram measured at $T = 4.7$~K $< T_{\rm N}$.
In this low temperature experiment, a serious background due to vacuum chamber walls of the refrigerator appears in a TOF region of $22.5 < t_{\rm TOF} < 31$~ms.
Therefore, the data in the range were removed in the figure.
It can be seen that the diffraction pattern below $T_{\rm N}$ is essentially the same as that at the room temperature.
Hence, no change in the crystal structure is concluded in the present diffraction experiment.

\begin{figure}
\includegraphics[width=\linewidth, angle=0]{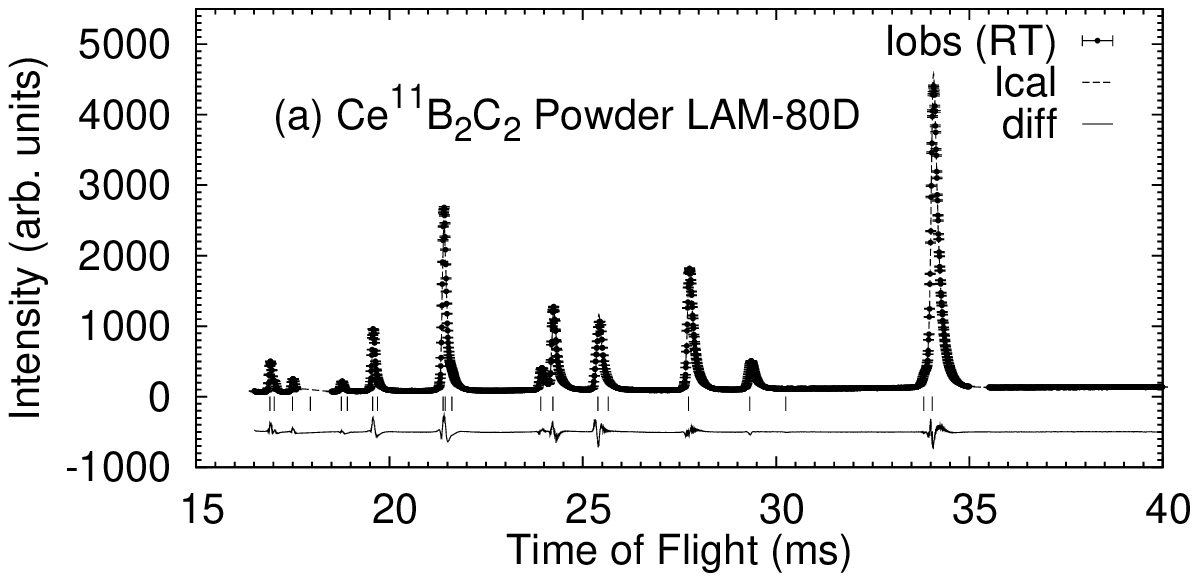}
\includegraphics[width=\linewidth, angle=0]{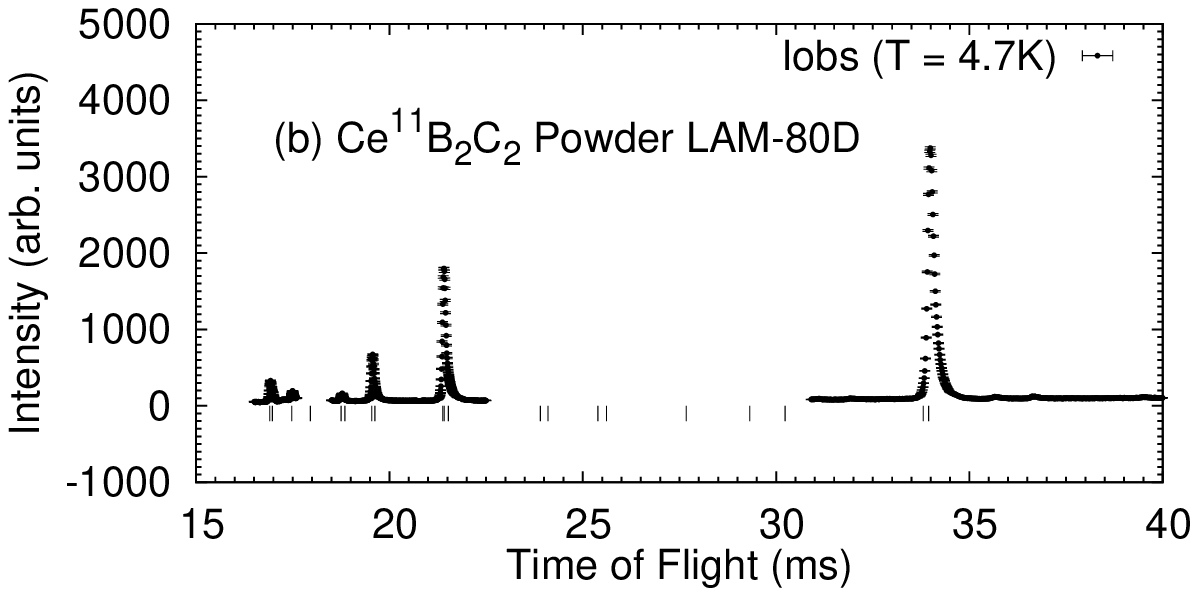}
\caption{(a) Powder diffraction patterns of Ce$^{11}$B$_{2}$C$_{2}$ at the room temperature.
The calculated diffraction pattern is shown in the figure by the dashed line, whereas the difference between the calculated and observed intensities is depicted by the solid line.
(b) Powder diffraction patterns of Ce$^{11}$B$_{2}$C$_{2}$ at $T = 4.7$~K.
The huge gap with no experimental data is due to the contamination of serious background due to the refrigerator chamber walls.}
\end{figure}

\begin{table}
  \begin{tabular}{cc}
    \hline    \hline
    Parameters\ \ & $T = $ RT\\ \hline
    $a$ (\AA) & 5.3948(3)\\
    $c$ (\AA) & 3.8646(3)\\
    $V$ (\AA$^3$) & 112.47(1)\\
    $x_{\rm B}$ & 0.3653(7)\\
    $x_{\rm C}$ & 0.1614(7)\\
    $B^{\rm iso}_{\rm Ce}$ & 5.4(3) \\
    $B^{\rm iso}_{\rm B}$ & 4.2(2) \\
    $B^{\rm iso}_{\rm C}$ & 3.5(2) \\
    $R_{\rm wp}$ & 0.0676\\
    \hline    \hline
  \end{tabular}
  \caption{
    Structural parameters for Ce$^{11}$B$_2$C$_2$ at the room temperature.
    The space group was assumed to be $P4/mbm$. The Ce atoms occupy the $2(a)$ sites at (0,0,0), whereas the B and C atoms occupy the $4(h)$ sites at $(x_{\rm B}, x_{\rm B} + 1/2, 1/2)$ and $(x_{\rm C}, x_{\rm C} + 1/2, 1/2)$, respectively ($Z = 2$).
  }
\end{table}

\subsection{Magnetic structure}

\begin{figure}
\includegraphics[width=\linewidth,angle=0]{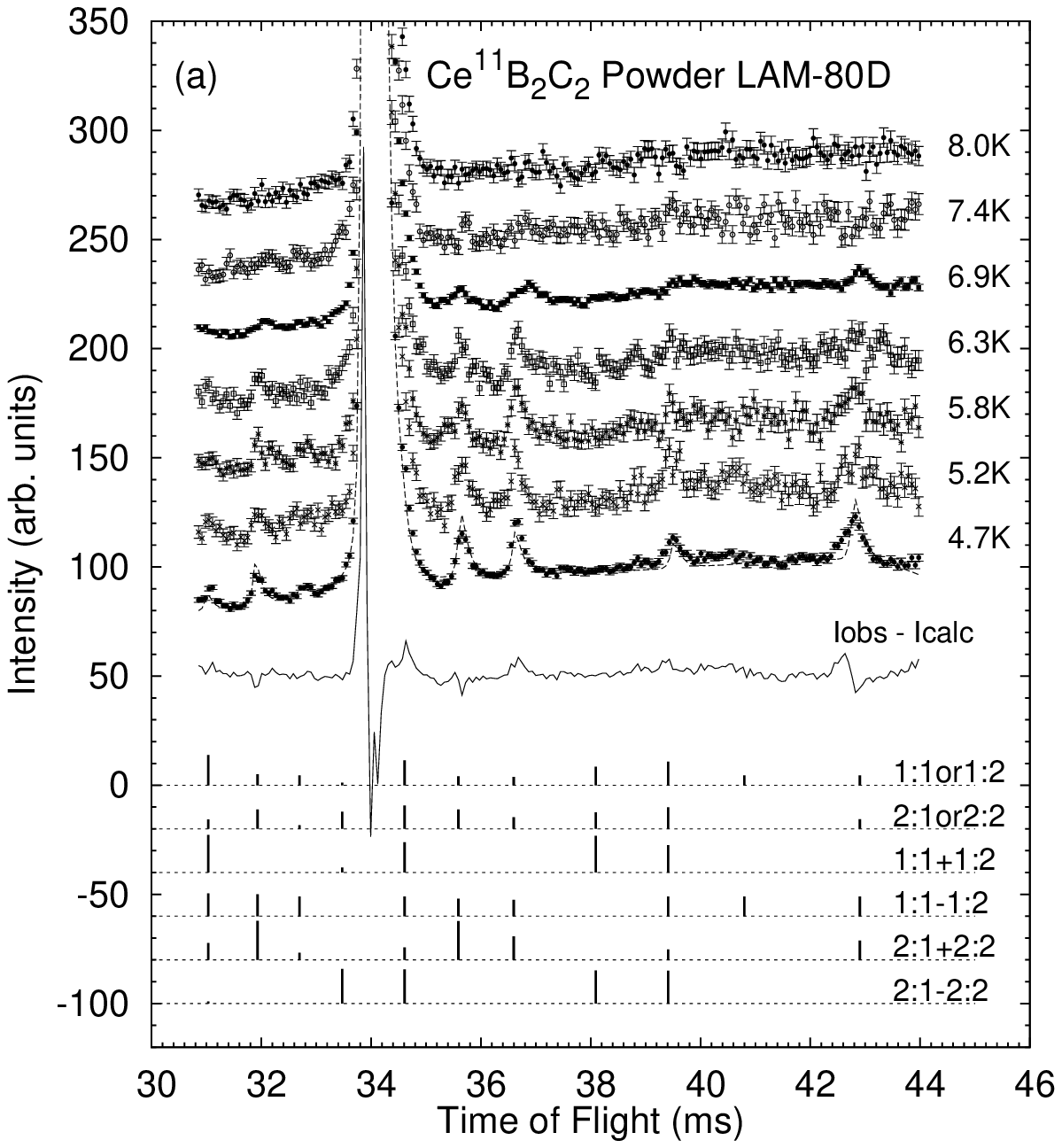}
\includegraphics[width=\linewidth,angle=0]{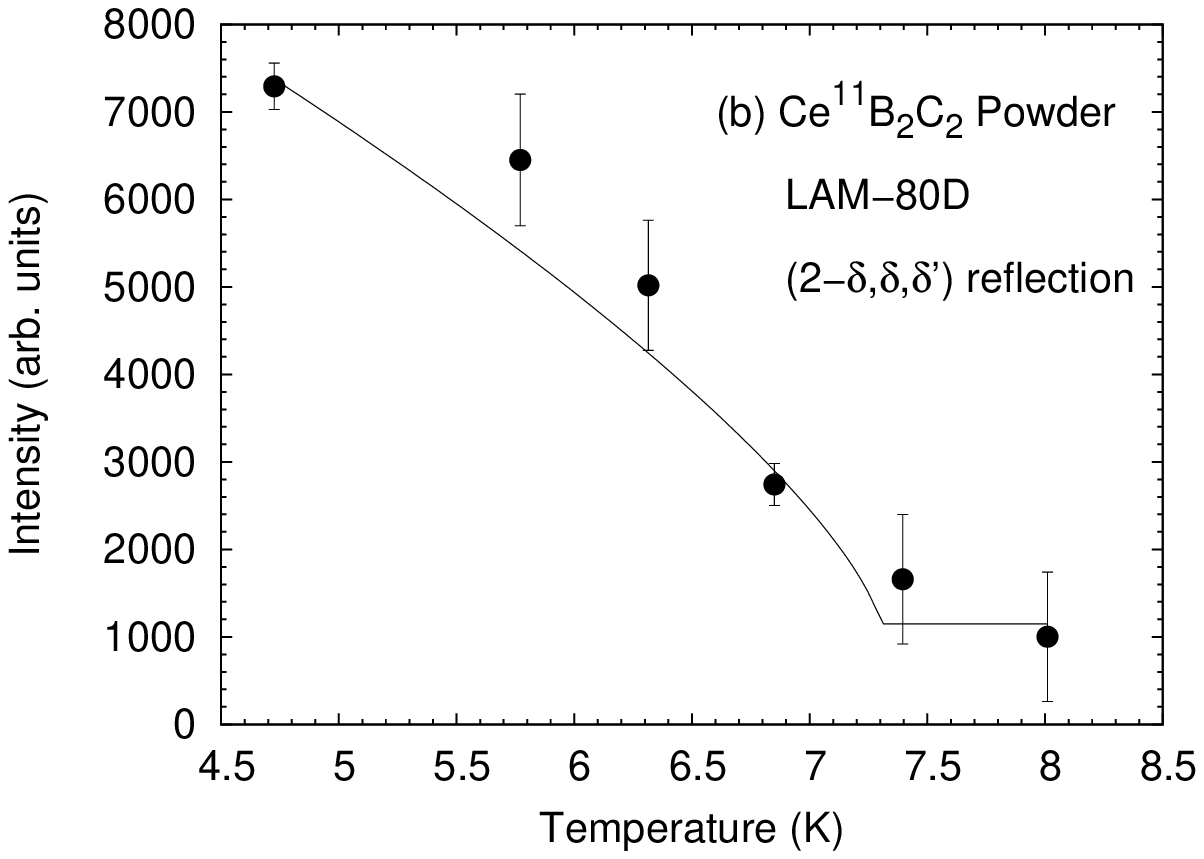}
\caption{(a) Neutron powder diffraction patterns at several temperatures between 8.0~K and 4.7~K.
Magnetic Bragg reflections clearly appear at the low temperatures.
Dashed line stands for the result of the Rietveld fitting for the pattern at 4.7~K, whereas the solid line denotes the difference between the observed and calculated intensities.
Vertical lines shown in the lower part of the figure are the magnetic Bragg peak positions and intensities for magnetic structures given by the linear combinations of the irreducible representations IR$n$:$m$.
Note that only IR2:1+IR2:2 can reproduce the observed diffraction pattern.
(b) Temperature dependence of the integrated intensity for the $(2-\delta, \delta, \delta')$ reflections, appearing at $t_{\rm TOF} \approx 36.7$~ms.
Solid line is a guide to the eyes.}
\end{figure}

Figure 2(a) shows the powder diffraction patterns at large TOF ($30 < t_{\rm TOF} < 45$~ms) in the low temperature range ($T < 8.0$~K).
Several additional reflections appear in the diffractogram below $T_{\rm N}$.
A huge Bragg peak at $t_{\rm TOF} = 33.9$~ms is the superposition of the nuclear 111 and 200 reflections.
In the diffractogram, one can clearly see evolution of several new Bragg peaks below $7.0$~K.
They are two orders of magnitude weaker than the nuclear Bragg reflections.
(Because of the weak intensity, they can hardly be recognized in the smaller TOF region.)
Figure~2(b) shows temperature dependence of the integrated intensity of the Bragg peak appearing at $t_{\rm TOF} \approx 36.7$~ms.
On cooling, the intensity starts to increase at $T \approx 7$~K, which is in good agreement with the macroscopically determined ordering temperature $T_{\rm N} = 7.3$~K.
This confirms the magnetic origin of the newly appearing Bragg peaks.
We also find that these magnetic Bragg peaks can be indexed as satellite peaks of the nuclear reflections using the previously reported magnetic modulation vector, ${\bm q} = (\delta, \delta, \delta')$, and its symmetrically equivalents. ($\delta$ and $\delta'$ are refined in the present study as described later.)
Thus, the present powder diffraction result is consistent with the previous single-crystal study, and provides intensity information for more Bragg reflections that is mandatory for the spin structure analysis.

Because of the limited TOF range and arbitrariness for magnetic reflection indexing due to powder averaging, it is not straightforward to find a spin structure model directly from the observed diffraction pattern.
Thus, to find possible spin-structure candidates, we use the magnetic representation analysis introduced by Izyumov {\it et al.}.~\cite{izy91}
In this method, structure candidates are given by linear combinations of magnetic basis vectors of the irreducible representations in the paramagnetic phase.
In the Landau theory of second order phase transition, a single irreducible representation may be selected as a symmetry of the ordered phase.
In reality, it frequently happens that two or more irreducible representations are necessary to reproduce the symmetry of the ordered phase.
Nonetheless, the number of the necessary representations is usually small.
Thus, we may expect that the ordered phase in CeB$_2$C$_2$ may be given by a combination of a few irreducible representations.
Here, we try to find the magnetic structure model using the smallest number of the magnetic basis vectors.
For this purpose, a representation analysis code, named {\footnotesize MBASE}, has been newly developed, which can calculate magnetic representation basis vectors for arbitrary ${\bm k}$-group and magnetic-ion positions.~\cite{sato07}

Assuming a single-${\bm q}$ structure with multiple domains of the equivalent modulation vectors, direction of a spin (or total angular momentum) at the $d$-th site in the $l$-th unit cell may be generally written as:
\begin{equation}
  \langle {\bm J}_{l,d}\rangle =  \frac{J}{2} [{\bm a}_{d} \exp(-{\rm i}{\bm q} \cdot {\bm R}_l) + {\bm a}_{d}^{*} \exp({\rm i}{\bm q} \cdot {\bm R}_l)],
\end{equation}
where ${\bm R}_l$ denotes the position of the $l$-th unit cell.
The polarization vector ${\bm a}_{d}$ is given as a linear combination of the magnetic basis vectors:
\begin{equation}
{\bm a}_{d} = \sum_{n,m} c_{n,m} {\bm a}_{n,m,d},
\end{equation}
where ${\bm a}_{n, m, d}$ denotes the basis vector of the irreducible representation IR$n:m$ for the spin at the $d$-th site.
The basis vectors for the ${\bm q}$-domain and for the domains with symmetrically equivalent modulations (arms) are listed in Table~II.
In the present Rietveld analysis, the domains with the equivalent modulations are assumed to be equally populated.
For the selection of the basis vectors, we note that there is very large anisotropy in the magnetic susceptibility; $\chi_{\rm c}$ is considerably smaller than $\chi_{\rm a}$ and $\chi_{110}$.~\cite{onodera00}
This indicates that spins most likely lie in the basal plane.
Hence, we may use only the in-plane basis vectors.
In the bottom part of Fig.~2(a), reflection positions and intensities from the magnetic structures given by single or linear combinations of IR$n:m$ are shown by the vertical thick solid lines.
We find that a single irreducible representation cannot reproduce the peak positions; for instance IR1:1 definitely gives a peak at $t_{\rm TOF} \approx 38.0$~ms, which cannot be found in the observed diffraction pattern.
Using two vectors with a constraint to the coefficient $c_{n,m} = \pm \sqrt{2}$, we find that the combination IR2:1 + IR2:2 [{\it i.e.}, ${\bm a}_{d} = ({\bm a}_{2,1,d} + {\bm a}_{2,2,d})/\sqrt{2}$], can satisfactorily reproduce the observed magnetic reflection positions.
This combination corresponds to a sinusoidally modulated structure with spin polarization parallel to the [110] direction.
The spin structure in the basal $c$-plane is schematically shown in Fig.~3.

\begin{table}
  \begin{tabular}{ccc}
    \hline\hline
    IR$n$:$m$ & ${\bm a}_{n,m, 1}$ & ${\bm a}_{n,m,2}$\\
    \hline
    ${\bm q} = (\delta, \delta, \delta')$ & & \\
    1:1 & $(1,0,0)$ & $(0,-\epsilon^{*},0)$ \\
    1:2 & $(0,1,0)$ & $(-\epsilon^{*},0,0)$ \\
    1:3 & $(0,0,1)$ & $(0,0,-\epsilon^{*})$ \\
    2:1 & $(1,0,0)$ & $(0,\epsilon^{*},0)$ \\
    2:2 & $(0,1,0)$ & $(\epsilon^{*},0,0)$\\
    2:3 & $(0,0,1)$ & $(0,0,\epsilon^{*})$\\
    \hline
    ${\bm q} = (-\delta, \delta, \delta')$ & & \\
    1:1 & $(1,0,0)$ & $(0,1,0)$\\
    1:2 & $(0,1,0)$ & $(1,0,0)$\\
    1:3 & $(0,0,1)$ & $(0,0,-1)$\\
    2:1 & $(1,0,0)$ & $(0,-1,0)$\\
    2:2 & $(0,1,0)$ & $(-1,0,0)$\\
    2:3 & $(0,0,1)$ & $(0,0,1)$\\
    \hline
    ${\bm q} = (\delta, -\delta, \delta')$ & & \\
    1:1 & $(1,0,0)$ & $(0,1,0)$ \\
    1:2 & $(0,1,0)$ & $(1,0,0)$ \\
    1:3 & $(0,0,1)$ & $(0,0,-1)$ \\
    2:1 & $(1,0,0)$ & $(0,-1,0)$\\
    2:2 & $(0,1,0)$ & $(-1,0,0)$ \\
    2:3 & $(0,0,1)$ & $(0,0,1)$ \\
    \hline
    ${\bm q} = (-\delta, -\delta, \delta')$ & & \\
    1:1 & $(1,0,0)$ & $(0,-\epsilon,0)$ \\
    1:2 & $(0,1,0)$ & $(-\epsilon,0,0)$ \\
    1:3 & $(0,0,1)$ & $(0,0,-\epsilon)$ \\
    2:1 & $(1,0,0)$ & $(0,\epsilon,0)$\\
    2:2 & $(0,1,0)$ & $(\epsilon,0,0)$\\
    2:3 & $(0,0,1)$ & $(0,0,\epsilon)$\\
    \hline\hline
    \end{tabular}
  \caption{Magnetic representation basis vectors ${\bm a}_{n,m,d}$ for the irreducible representations IR$n:m$.
	They are calculated for the four magnetic domains with the modulation vectors ${\bm q} = (\delta, \delta, \delta'), (-\delta, \delta, \delta'), (\delta, -\delta, \delta')$, and $(-\delta, -\delta, \delta')$.
    $\epsilon$ is defined as $\epsilon = \exp(2 \pi {\rm i} \delta)$.
    The basis vectors for $-{\bm q}$ are given by complex conjugates of the above vectors.
    The vector ${\bm a}_{n,m,1}$ is for the (0,0,0) site, whereas ${\bm a}_{n,m,2}$ for the $(\frac12,\frac12,0)$ site.
}
\end{table}

\begin{figure}
\includegraphics[width=\linewidth,angle=0]{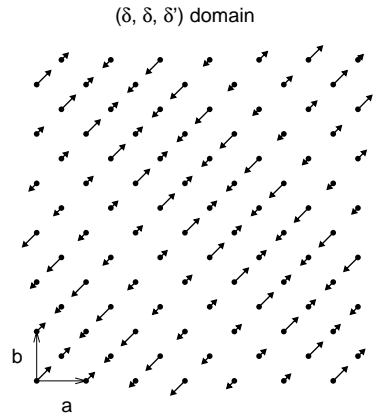}
\caption{Schematic drawing of the spin structure in the basal $c$-plane at $T = 4.7$~K determined in the present study.
Only one domain with ${\bm q} = (\delta, \delta, \delta')$ is shown in the figure.}
\end{figure}

The profile fitting has been performed assuming the IR2:1+IR2:2 structure; the fitting parameters were the modulation vector ${\bm q}$ and amplitude $J$.
Dashed line in the Fig.~2(a) shows the result of the profile fitting to the diffraction pattern taken at 4.7~K.
The difference between the calculation and the observation is also shown in the figure by the solid line.
Reasonable coincidence can be found in the figure between the calculated and observed intensities, despite the rather deficient statistics of the experimental data.
From the profile fitting, $J$ is estimated as $J = 1.8(2)$, corresponding to the maximum magnetic moment of $g_J J \mu_{\rm B}= 1.5(2)$~$\mu_{\rm B}$ at 4.7~K.
The ${\bm q}$-vector is refined as ${\bm q} = [0.167(3), 0.167(3), 0.114(3)]$.
It may be noteworthy that this ${\bm q}$-vector is quite close to the commensurate value of $(1/6, 1/6, 1/9)$, although this cannot be concluded in the present powder diffraction measurements; a single-crystal experiment is highly desired to conclude this issue.

\begin{figure}
\includegraphics[width=\linewidth,angle=0]{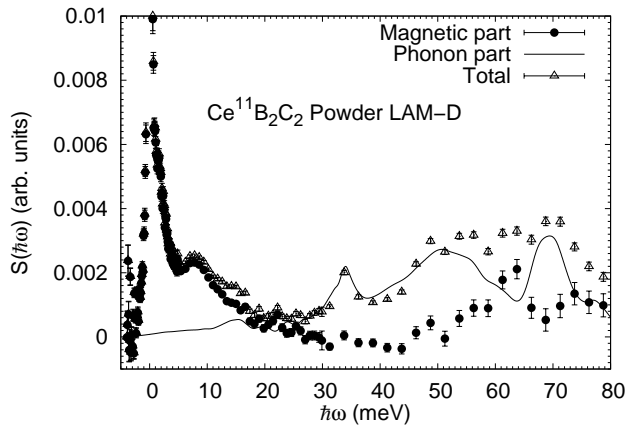}
\caption{Neutron inelastic scattering spectrum at $T = 15.5$~K. 
The open triangles and closed circles stand for the total scattering intensity and the phonon-subtracted magnetic intensity, respectively.
Phonon contribution estimated from the $T = 100$~K data is shown by the solid line. 
}
\end{figure}

\subsection{Inelastic scattering}
\begin{figure}
\includegraphics[width=\linewidth,angle=0]{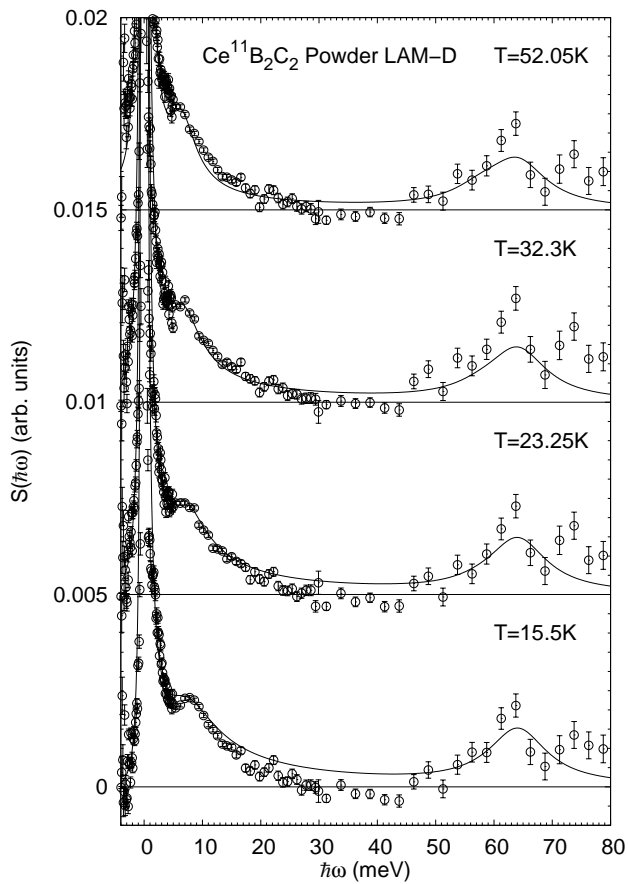}
\caption{Magnetic scattering spectra at $T = 52.05, 32.3, 23.25$, and 15.5~K (from top to bottom).
The solid lines stand for the fitting to the model scattering function Eq.~(\ref{eq:sqwWideE}), combined with a delta-function like elastic peak.
See text for details.
}
\end{figure}
\begin{figure}
\includegraphics[width=\linewidth,angle=0]{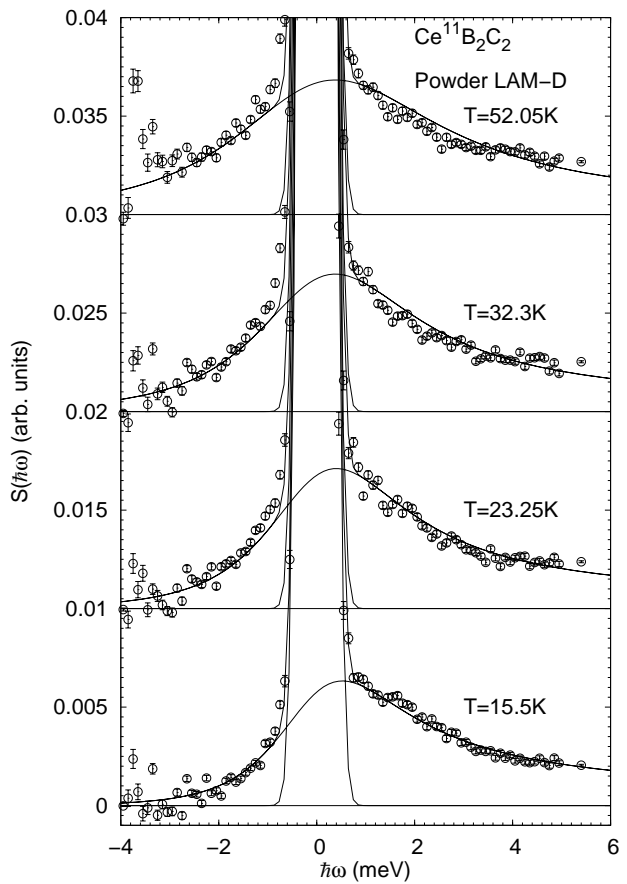}
\caption{Magnetic scattering spectra at $T = 52.05, 32.3, 23.25$, and 15.5~K (from top to bottom).
The solid lines stand for the fitting to the pseudo-Voigt quasielastic signal, and a delta-function like elastic peak.
See text for details.
}
\end{figure}

\subsubsection{Phonon subtraction procedure}

Next, to investigate the spin fluctuations and excitations, we have performed the inelastic scattering experiment using the same powder sample.
To reliably obtain magnetic scattering intensity from the raw inelastic spectrum, a phonon contribution has to be carefully removed.
Hence, we first make an estimation of the phonon contribution using high temperature data at $T = 100$~K.
For the rare-earth compounds in a paramagnetic temperature range, magnetic scattering may originate from local (single-site) transitions, and thus, $Q$-dependence of the magnetic scattering may be dominantly given by the magnetic form factor.
On the other hand, the phonon scattering may approximately exhibit $Q^2$ dependence.
Thus, we assume the following $Q$ dependence for the total ({\it i.e.}, magnetic + phonon) scattering function:
\begin{equation}\label{eq:phonon}
S_{\rm tot}(Q, \hbar \omega) = A_{\rm mag}(\hbar \omega)[f_{\rm mag}(Q)]^2 + A_{\rm ph}(\hbar \omega)Q^2,
\end{equation}
where $A_{\rm mag}(\hbar \omega)$ and $A_{\rm ph}(\hbar \omega)$ stand for the $Q$ independent parts of the magnetic and phonon scattering, whereas $f_{\rm mag}(Q)$ for the magnetic form factor of the Ce$^{3+}$ ions.~\cite{fre79}
By comparing the inelastic spectra at $T = 100$~K measured with the two different scattering angles ($2\theta = 35^{\circ}$ and $85^{\circ}$), we estimate the phonon contribution $A_{\rm ph}(\hbar \omega)Q^2$.
Then, the phonon contribution at low temperatures is obtained using the $[1-\exp(-\hbar\omega/k_{\rm B}T)]^{-1}$ dependence.
Figure~4 exemplifies this phonon subtraction procedure for the representative data at $T = 15.5$~K.
The solid line stands for the estimated phonon intensity, whereas phonon-subtracted magnetic intensity is shown by the filled circles.
In the low energy region, the phonon intensity is considerably small, whereas the phonon intensity becomes dominant in the higher energy region ($\hbar \omega > 30$~meV).
This large phonon contamination reduces statistical accuracy of the estimated magnetic intensity, however, one finds that there definitely exists finite magnetic intensity for $\hbar \omega > 50$~meV.

\subsubsection{Inelastic spectrum in the paramagnetic phase ($T > T_{\rm N}$)}

The phonon-subtracted magnetic scattering spectra at four representative temperatures $T = 52.05, 32.3, 23.35$, and 15.5~K in the paramagnetic phase are shown in Figs.~5 and 6.
Three peaks are recognized in the spectra at all the temperatures, in addition to the sharp $\delta$-function like elastic peak: (i) quasielastic component centered at $\hbar \omega = 0$ (Fig.~6); (ii) inelastic peak at $\hbar \omega \approx 8$~meV (Fig.~5); (iii) weak inelastic peak at $\hbar \omega \approx 65$~meV (Fig.~5).
Since those spectra were measured in the paramagnetic phase, the peaks are most likely due to transitions between the CEF splitting levels.
Hence, the observed spectra are analyzed using the following CEF Hamiltonian derived for Ce$^{3+}$ in the $C_{4h}$ site symmetry of the $2(a)$ site in CeB$_2$C$_2$:
\begin{equation}
H_{\rm CEF} = B_{20}O_{20} + B_{40} O_{40} + B_{44} O_{44} + B_{4-4} O_{4-4},
\end{equation}
where $O_{nm}$ stands for the Stevens operators.~\cite{hutchings64}
Under above CEF, $J = 5/2$ multiplet of Ce$^{3+}$ splits into three Kramers' doublets.
The transition strengths between the CEF splitting levels are given as follows:
\begin{eqnarray}\label{eq:CEF}
b^{\alpha}_{nm} &=& \frac{2{\rm e}^{-E_n/k_{\rm B}T}}{Z} 
\frac{|\langle n |J^{\alpha}| m \rangle|^2}{E_m - E_n},\ \ \ (m \neq n)\nonumber\\
b^{\alpha}_{nn} &=& \frac{{\rm e}^{-E_n/k_{\rm B}T}}{Z} 
\frac{|\langle n |J^{\alpha}| n \rangle|^2}{k_{\rm B}T},\ \ \ \mbox{(otherwise)}
\end{eqnarray}
where $|n\rangle$ and $|m\rangle$ are wave functions for the initial and final states of the CEF splitting levels (see Fig.~7 for numbering of the states), and $Z$ is the partition function.
The scattering function from a powder sample may be given by a sum of spectral weights of the CEF transitions:
\begin{widetext}
\begin{equation}\label{eq:sqwWideE}
S(Q,\hbar \omega)_{\rm inel} = \frac{2}{3}\left [ \frac{1}{2} g_J f_{\rm mag}(Q) \right ]^2 \frac{N \hbar \omega}{1 - \exp(-\hbar \omega/k_{\rm B} T)}\sum_{nm\alpha}b^{\alpha}_{nm} P_{nm}(\hbar \omega; \hbar\omega_{nm},\Gamma_{nm}).
\end{equation}
In the above equation, we assume a Lorentzian-type peak profile for the inelastic CEF excitations:
\begin{equation}
P_{nm}(\hbar\omega; \hbar\omega_{nm}, \Gamma_{nm}) = \frac{\Gamma_{nm}}{\pi} \left [ \frac{1}{4(\hbar \omega - \hbar\omega_{nm})^2 + \Gamma_{nm}^2} + \frac{1}{4(\hbar \omega + \hbar\omega_{nm})^2 + \Gamma_{nm}^2} \right ].
\end{equation}
\end{widetext}
For the quasielastic peak shape, we assume a pseudo-Voigt function, which is a reasonable approximation of Lorentzian function convoluted by a Gaussian-shaped instrumental resolution function.~\cite{thompson87,wertheim74}
For the pseudo-Voigt function, the width of the unconvoluted Lorentzian is denoted by $\Gamma_{nn}$ (FWHM).
We note that the lowest order coefficient $B_{20}$ can be determined precisely using a single crystal magnetization measurement; $B_{20}$ is estimated as 6.34~K in the earlier work.~\cite{onodera00}
Thus, we fix $B_{20}$, and try to find the optimum values for $B_{40}$, $B_{44}$ and $B_{4-4}$ that reproduce all the inelastic spectra in the paramagnetic phase simultaneously.
In the fitting, we assume that temperature dependence of the Hamiltonian parameters $B_{20}, B_{40}, B_{44}$, and $B_{4-4}$ is negligible in the present temperature range.
On the other hand, most of the width parameters $\Gamma_{nm}$ are assumed as temperature dependent; only $\Gamma_{13}$ and $\Gamma_{23}$ have to be fixed to the empirical value 12~meV because of the insufficient statistics in the high energy regions.
The resulting optimum parameters for the CEF Hamiltonian are $B_{40} = -1.84(3)$~K, $B_{44} = 10.1(2)$~K and $B_{4-4} = -2.8(6)$~K.
Calculated spectra using these parameters are shown in Fig.~5 by the solid lines.
Reasonable coincidence can be seen between the observed and calculated intensities at all the temperatures.
The fact that the temperature dependence is well reproduced confirms the CEF (magnetic) origin of the three peaks.
It should be noted here that parameter $B_{4-4}$ could not be estimated reliably; its uncertainty range is considerably larger than those of other parameters.
This parameter exists in the CEF Hamiltonian because of the broken four fold symmetry due to the B-C ordering.
Since B and C have relatively similar electronegativity, the symmetry breaking may possibly be moderate, and thus we might infer that $B_{4-4}$ may be irrelevant.

Earlier neutron inelastic scattering experiment provides considerably different Hamiltonian parameters; $B_{40} = 0.024(2)$~meV ($= 0.28$~K) and $B_{44} = 0.356(1)$~meV ($= 4.13$~K).~\cite{hillier06}
The discrepancy is due to the assignment of the highest inelastic peak; the previous study assigned the highest energy peak to a very broad hump found at $\hbar \omega \approx 23$~meV in their spectrum.
The hump is completely absent in the presently observed inelastic spectrum.
It may be noteworthy that a weak peak in phonon density of states of the elemental aluminum, which is commonly used for sample cells in inelastic neutron scattering, exists in this energy range, and thus this may become a source of uncertainty.
On the other hand, we clearly see the highest energy peak at $\hbar \omega \approx 65$~meV, which is out of the observation energy range of the previous study.
We also note that earlier macroscopic study provides rough estimate of the CEF levels at 102~K and 1110~K,~\cite{ishimoto00} supporting our result.

\begin{figure}
\includegraphics[width=\linewidth,angle=0]{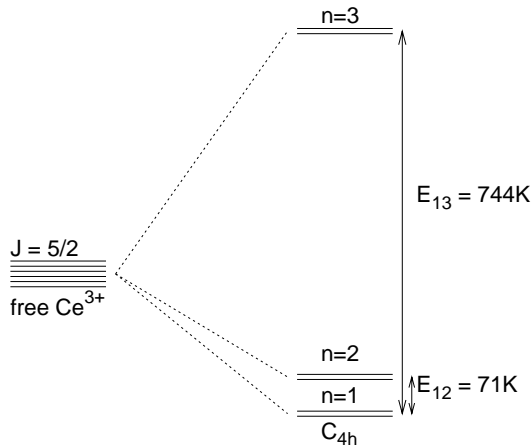}
\caption{Schematic illustration of the CEF splitting scheme in the CeB$_2$C$_2$.
The energy separations of the Kramers' doublets are determined in the present study.
See text for details.
}
\end{figure}

Among the three peaks, the quasielastic peak is of particular interest in the Ce compounds, since it reflects the low-energy scattering process between the ground state doublets and the conduction electrons.
Hence, we parameterize the quasielastic-peak width as a function of temperature.
To obtain the width precisely, the fitting is performed in the limited energy range $-3 < \hbar \omega < 6$~meV using the pseudo-Voigt function with the unconvoluted Lorentzian width $\Gamma_{\rm qel}$ as fitting parameter.
The fitting result is depicted by the solid lines in Fig.~6.
Obtained temperature variation of the quasielastic peak width $\Gamma_{\rm qel}$ is shown in Fig.~8.
The quasielastic peak becomes narrower on cooling, as generally seen in heavy-fermion systems.
It may be noteworthy that the quasielastic width is noticeably large even at $T \approx 12$~K, just above $T_{\rm N}$.
This indicates that quasielastic fluctuations due to the scattering by the conduction electrons are still dominant in vicinity of the ordering temperature.

\begin{figure}
\includegraphics[width=\linewidth,angle=0]{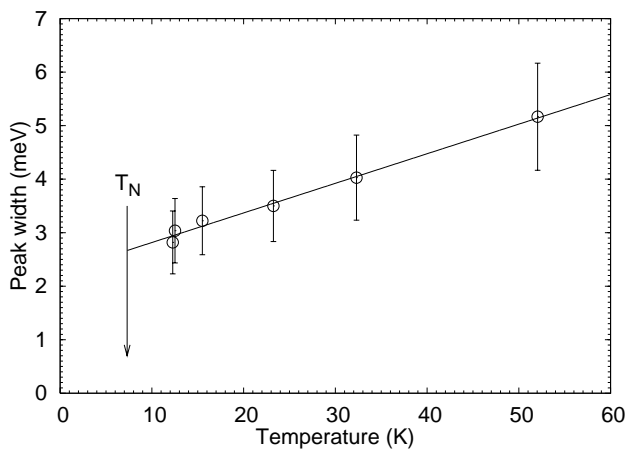}
\caption{Temperature dependence of the quasielastic peak width $\Gamma_{\rm qel}$ (FWHM) in the paramagnetic temperature range.
The solid line is a guide to the eyes.}
\end{figure}

\subsubsection{Inelastic spectrum in the ordered phase ($T < T_{\rm N}$)}

\begin{figure}
\includegraphics[width=\linewidth,angle=0]{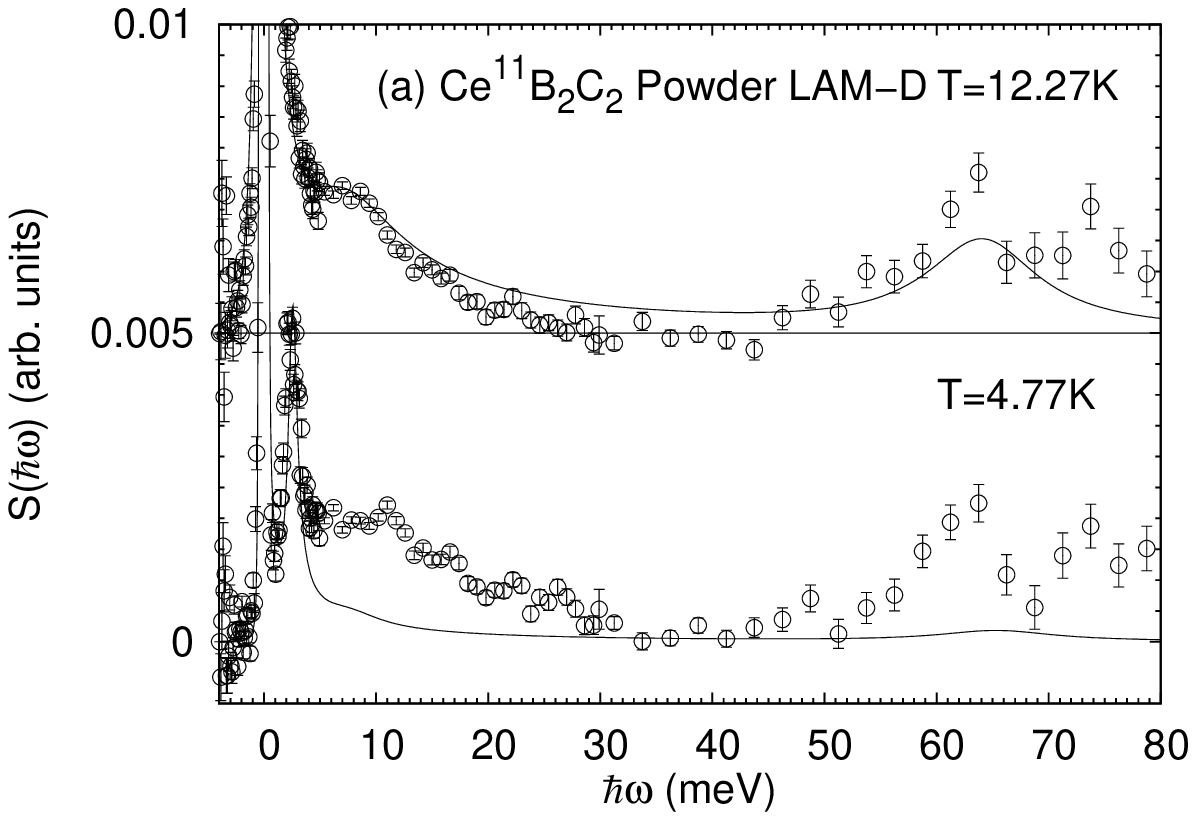}
\includegraphics[width=\linewidth,angle=0]{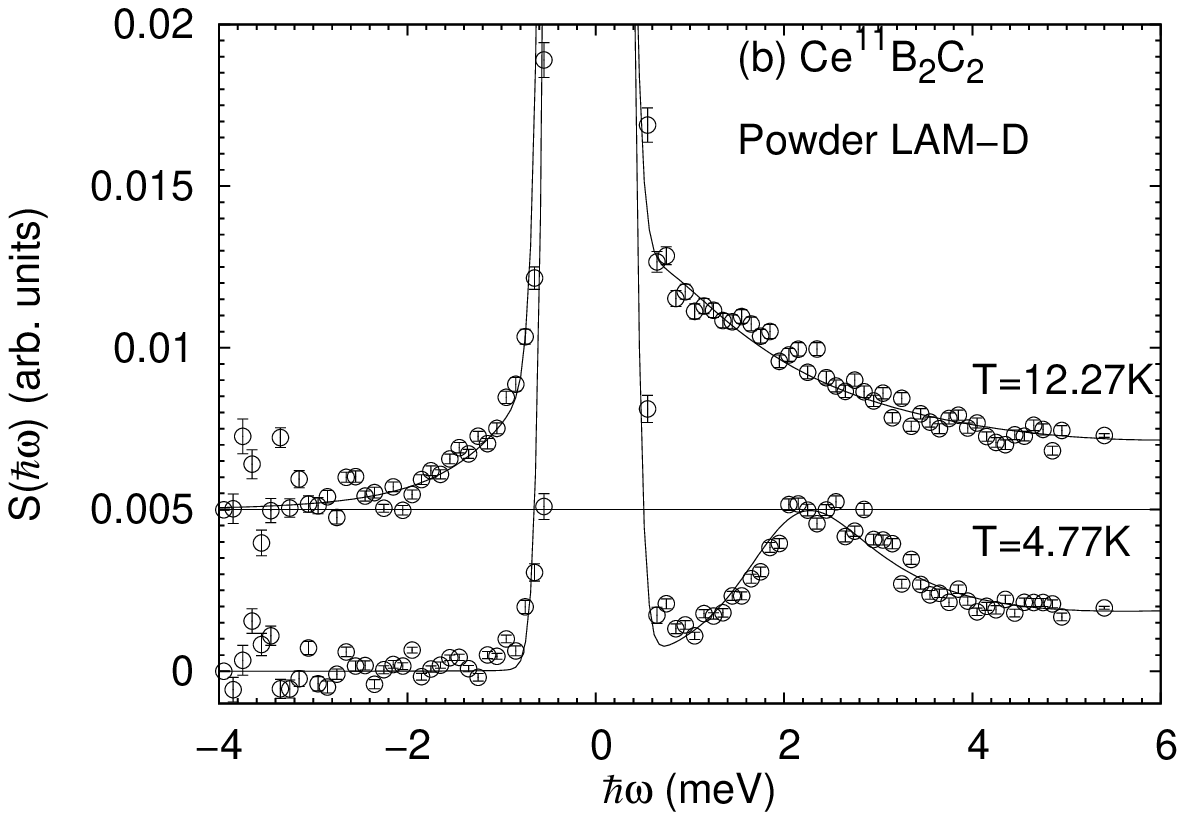}
\caption{(a) Neutron inelastic spectra at $T = 12.27$~K $> T_{\rm N}$ and $T = 4.77$~K $< T_{\rm N}$. 
Solid line is the calculated spectrum assuming the CEF Hamiltonian Eq.~(\ref{eq:CEF}), whereas the dashed line for $T = 4.77$~K is the result of the scattering-function calculation assuming an additional Zeeman term in the CEF Hamiltonian.
(b) Excitation spectra in the low energy region.  Solid lines in this plot are guides to the eyes.
}
\end{figure}

As seen in the previous section, the temperature dependence of inelastic spectrum is found to be moderate in the paramagnetic phase.
However, it shows drastic change across $T_{\rm N}$.
Representative inelastic spectrum below $T_{\rm N}$ is shown in Fig.~9, in comparison with the paramagnetic spectrum at $T = 12.27$~K.
The broad quasielastic signal around $\hbar\omega = 0$ disappears below $T_{\rm N}$, and a new inelastic peak develops at $\hbar \omega = 2.1$~meV, instead.
Note that the newly appearing 2.1~meV peak is quite sharper than that of the quasielastic peak at $T > T_{\rm N}$.
Thus, the low-energy spin fluctuations in the ground state doublet are strongly suppressed on ordering.
This strongly suggests dissociation of $4f$ electrons from the conduction electrons, {\it i.e.} formation of the localized moment in the ordered phase.

On the origin of the sharp inelastic peak, several possibility may be anticipated.
The simplest possibility may be splitting of the ground state doublet by the internal (exchange) molecular field appearing in the ordered phase.
To check this possibility, we calculated the single-site CEF excitation spectrum under the internal field by introducing the Zeeman term $H_{\rm Zeeman} = g_{J}\mu_{\rm B}{\bm J} \cdot {\bm H}_{\rm int}$ in the CEF Hamiltonian Eq.~(\ref{eq:CEF}).
Direction of the internal field ${\bm H}_{\rm int}$ is parallel to the spin direction [110].
The calculated inelastic spectrum assuming $H_{\rm int} = 17$~T is shown in Fig.~9(a) by the dotted line.
The peak position is reproduced by introducing the internal field, however, relative spectral weights of the peaks are apparently inconsistent with the observed spectrum.
Since in the sinusoidally modulated structure the molecular field is not uniform, another possibility may be the doublet splitting due to the distributed (non-uniform) internal fields.
However, the distribution will apparently introduce peak broadening, and thus will not reproduce the sharpness of the 2.1~meV peak.
As above, the single-site CEF origin is unlikely for the sharp inelastic peak.
As another possibility, we may thus speculate that the inelastic peak plausibly stems from a certain collective excitation of the interacting Ising-like spins formed by the ground state doublets.
To pursue the origin of the 2.1~meV inelastic peak, further study on its ${\bm Q}$-dependence using a single crystal is necessary.
Such a single-crystal inelastic scattering experiment is in progress.

It should be reminded that the long-period sinusoidally modulated structure has a number of spins that have strongly reduced average spin magnitudes $\langle S \rangle << S$ (see Fig.~3).
In a localized spin system, such reduced spins are realized by thermal fluctuations, and thus becomes unstable for further lowering temperature.~\cite{sato94}
On the other hand, for an incommensurate phase in heavy-fermion systems, such reduced spins are formed by quantum fluctuations; coupling with conduction electrons enables a formation of spin-density-wave-type (SDW) order, as typically seen in Ce$_x$La$_{1-x}$Ru$_2$Si$_2$~\cite{raymond01} or Ce(Ru$_{1-x}$Rh$_x$)$_2$Si$_2$.~\cite{kawarazaki97}
In both the cases, either thermal or quantum spin fluctuations should remain in the sinusoidally modulated phase.
In contrast to the above understanding of the sinusoidally modulated phase, quasielastic fluctuations are completely suppressed in CeB$_2$C$_2$.
This is a very unique characteristic of the sinusoidally modulated phase in the CeB$_2$C$_2$, and further experimental as well as theoretical study is highly desired to clarify this issue.

Finally, we compare the ordered spin structure of CeB$_2$C$_2$ with those in the heavy RE systems.
Incommensurately modulate structures have been frequently observed in the heavy RE systems, such as HoB$_2$C$_2$,~\cite{tobo01,ohoyama00} TbB$_2$C$_2$,~\cite{kaneko01,kaneko01b} and ErB$_2$C$_2$.~\cite{effantin85}
In HoB$_2$C$_2$, incommensurately modulated spin structure is realized blow $T_{\rm c1} = 5.9$~K with the two modulation vectors ${\bm q} = (1,0,0)$ and $(1 \pm \delta, \delta, \delta')$ where $\delta = 0.112$ and $\delta' = 0.04$.
Quadrupolar order is established at lower temperatures $T < T_{\rm C2} = 5.0$~K, where spin structure becomes commensurate with the four modulation vectors ${\bm q} = (1,0,0)$, $(0,1,1/2)$, $(0,0,0)$ and $(0,0,1/2)$.
Note that the spin modulation vectors in the quadrupolar ordered phase are identical to those in DyB$_2$C$_2$.
TbB$_2$C$_2$ also shows similar multi-${\bm q}$ incommensurate structure.
This compound does not show quadrupolar ordering in the zero external field, however, quadrupolar ordering is known to take place in a very low external magnetic field of 1~T, indicating that a likely situation is realized for the formation of the quadrupolar moment in the CEF ground state.
In contrast, ErB$_2$C$_2$ exhibits the single-${\bm q}$ sinusoidally modulated structure below $T_{\rm N} = 15.9$~K with the incommensurate modulation vector $q = (1+\delta, \delta, 0)$ ($\delta = 0.112$).
It is followed by a lower-temperature lock-in transition to ${\bm q} = (1,0,0)$ at $T_{\rm t} = 13.0$~K.
Such a lock-in transition is typical for the linearly polarized incommensurate phase in localized spin systems.
No quadrupolar ordering behavior was reported in this system.
In the present study, we have also observed a single-${\bm q}$ sinusoidal phase in the CeB$_2$C$_2$ with no quadrupolar degree of freedom.
Therefore, this result, in addition to the above observations for the other RE systems, supports the claim that the magnetic sector of the REB$_2$C$_2$ compounds has a tendency to form a long-period single-${\bm q}$ structure, and it is the effect of the quadrupolar degree of freedom that realizes the complex multi-${\bm q}$ structures.

\section{Conclusions}
We have performed neutron powder diffraction and inelastic scattering experiments on the cerium diborocarbide Ce$^{11}$B$_2$C$_2$.
In the powder diffraction study, we have clearly detected the magnetic reflections below the ordering temperature $T_{\rm N} = 7.3$~K.
With the aide of the magnetic representation analysis, we have determined the magnetic structure in the ordered phase as the sinusoidally modulated structure with the modulation vector ${\bm q} = [0.167(3), 0.167(3), 0.114(3)]$.
The inelastic study detects three magnetic peaks in the paramagnetic phase; the quasielastic signal around $\hbar \omega = 0$ and broad inelastic peaks at $\hbar \omega \approx 8$ and $65$~meV.
Upon cooling to $T < T_{\rm N}$, drastic change has been seen in the low energy spin fluctuation spectrum; a sharp inelastic peak develops at $\hbar \omega = 2.1$~meV.
This indicates that the spin fluctuations due to hybridization between the $4f$ and conduction electrons in the paramagnetic temperatures are strongly suppressed in the ordered phase.

\begin{acknowledgments}
The present authors thank Drs. I.~Tsukushi and K.~Shibata for the installation of the new detector bank at LAM-80ET, and Drs. H.~Kadowaki, K. Ohoyama and H. Onodera for valuable discussions.
This work was partly supported by a Grant-in-Aid for Encouragement of Young Scientists (B) (No. 16760537) and by a Grant-in-Aid for Creative Scientific Research (No. 16GS0417) from the Ministry of Education, Culture, Sports, Science and Technology of Japan.
\end{acknowledgments}


\end{document}